\documentclass[conference]{IEEEtran}
\ifCLASSINFOpdf
\else
\fi

\usepackage{mathtools}
\usepackage{standalone}
\usepackage{booktabs, dcolumn}

\usepackage{lscape}
\usepackage{rotating}
\usepackage{latexsym}
\usepackage{dcolumn}
\usepackage{booktabs}
\usepackage{amsmath}
\usepackage[algo2e,ruled,linesnumbered,boxed,commentsnumbered]{algorithm2e}
\usepackage{multirow}
\usepackage{graphicx}
\usepackage{amssymb}
\usepackage{pifont}
\newcommand{\cmark}{\ding{51}}%

\usepackage{amsmath,array,graphicx}
\usepackage{kantlipsum}

\usepackage[varg]{txfonts}
\usepackage[usestackEOL]{stackengine}

\usepackage{tabularx}

\usepackage[para]{threeparttable}
\usepackage{etoolbox}
\usepackage{lipsum}

\begin{document}
%
\title{A Comprehensive Survey of \\Ontology Summarization: Measures and Methods 
}



\author{
    \IEEEauthorblockN{Seyedamin Pouriyeh\IEEEauthorrefmark{1}, Mehdi Allahyari\IEEEauthorrefmark{2}, Krys Kochut\IEEEauthorrefmark{1},  Hamid Reza Arabnia\IEEEauthorrefmark{1}}
    \IEEEauthorblockA{\IEEEauthorrefmark{1}Computer Science Department, University of Georgia, Athens, GA, USA
    \\\{pouriyeh, kkochut, hra\}@uga.edu}
    \IEEEauthorblockA{\IEEEauthorrefmark{2}Computer Science Department, Georgia Southern University, Statesboro, USA
    \\mallahyari@georgiasouthern.edu}

}


%


\maketitle

\begin{abstract}
The Semantic Web is becoming a large scale framework that enables data to be published, shared, and reused in the form of \emph{ontologies}. Ontology which is considered as basic building block of semantic web consists of two layers including data and schema layer. With the current exponential development of ontologies in both data size and complexity of schemas, ontology understanding which is playing an important role in different tasks such as ontology engineering, ontology learning, etc., is becoming more difficult. Ontology summarization as a way to distill knowledge from an ontology and generate an abridge version to facilitate a better understanding is getting more attention recently. There are various approaches available for ontology summarization which are focusing on different measures in order to produce a proper summary for a given ontology.  In this paper, we mainly focus on the common metrics which are using for ontology summarization and meet the state-of-the-art in ontology summarization.  

\end{abstract}


%
\IEEEpeerreviewmaketitle

\section{Introduction}
\label{S:1}

In the recent years, we are facing an exponential  growth of the Semantic Web resources contains large amount of data such as million of semantic documents and billions of triples. Semantic Web is providing a large scale structure that facilitates data to be published, shared, and reused across different applications in the form of \emph
{ontology} \cite{brank2005survey}. Ontologies which are playing an important role in the deployment and development of the Semantic Web are usually represented  by two different layers forming different graphs. The first layer, called the schema layer or \emph{Resource Description Framework Schema (RDFS)}, functions as meta-data and describes the fundamental aspects of the data layer. The other layer, called the data layer or \emph{Resource Description Framework (RDF)}, stores the actual data according to the defined schema layer. 
The two layers make up a framework to represent knowledge bases, including concepts, entities, and relationships among them \cite{hlomani2014approaches}. This framework has been recognized as an important tool for publishing large volume of linked data that facilitate retrieving abundant knowledge. With the dramatic growth in both data size and schema complexity of ontologies, comprehending, exploring, and exploiting of those ontologies are becoming more difficult. Summarization in both the data and schema layer in order to receive an overview of data source is one possible solution that are getting more attention recently. In fact, providing an overview for a better understanding of an ontology can facilitate and reduce the cost of the next task(s) in various applications such as querying a complex data source \cite{ehrig2004ontology,davies2008semantic}, integrating different data sources \cite{trippe2017vision}, entity summarization \cite{pouriyeh2017lda}, labeling \cite{allahyari2017knowledge}, knowledge based summarization \cite{allahyari2017text}  and schema matching process \cite{bellahsene2011schema}.

In literature ontology summarization is defined as a technique of distilling knowledge from an ontology in order to produce an abridged version for different tasks \cite{zhang2007ontology}.Usually, the abridged version of an ontology covers the important nodes including entities or concepts.
Depending on applying various important measures to find the key nodes (entities or concepts) within an ontology, different ontology summarization techniques have been proposed. In this paper, we investigate the different available methods for ontology summarization task in both data and schema layer.

\section{Ontology Summarization }
As the size and the complexity of ontologies increase, there is a high demand in order to facilitate ontology understanding and help users take advantage of an ontology quickly. There are ontology management \cite{hepp2007ontology} techniques that aim to reduce the size and complexity of an ontology such as ontology partitioning \cite{d2007ontology}, ontology segmentation \cite{seidenberg2006web}  , and ontology modularization \cite{stuckenschmidt2009modular} but, they do not keep the most important information. Differently, ontology summarization attempt to provide efficient and effective models to grab knowledge from an ontology while preserving informative nodes. The expected outcome of ontology summarization, usually, is defined as sub-graphs or key nodes at the data and schema level that represents the most important nodes (entities or concepts). 

\section{Assessment Measures}
\label{assess}
A broad range of node importance meaning in the context of ontology has led to the emergence of many different algorithms that aim to highlight the most descriptive concepts and/or entities at the schema and data level respectively. 
Although, several measures have been proposed in this area but there is no generalized approach to extract the representative summary for an ontology. The main reason behind this scenario is that the application and the domain of the summary provide a guideline in order to select a proper set of measures. 
In this section we review the most popular measures that are using to extract the important node(s) within an ontology. We will consider an ontology $(O)$ in a form of a graph $G$ consists of a set of vertices ($V$) and edges ($E$) , $G = (V, E) $, and $N$ is the number of nodes in the graph $G$. Also, the following measures are applicable for nodes $(v, u, s, t) \in V$.

\subsection{Degree Centrality}
\label{S:2}
Degree centrality is simple measurement to calculate the importance of vertices in a graph  \cite{erkan2004lexrank,hoser2006semantic,zhang2007ontology}. The degree centrality of a node is a measure of local centrality of that node and it is determined through the sum of the edges that node has. This scenario is applicable for undirected graph (equation \ref{eq:dc}). For a directed graph, the degree centrality measure is divided into two categories including \emph{in-degree centrality} and \emph{out-degree centrality}. In-degree  and out-degree centrality are measured by counting the number of incoming and outgoing link from a particular node (equation \ref{eq:dcinout}). Nodes with higher degree centrality (In/out-degree centrality) are usually considered as more important nodes.

\begin{equation}
\label{eq:dc}
DC(v) =  | Number of edges(v) |
\end{equation}
\begin{equation}
\label{eq:dcinout}
DC_{in/out}(v) =  | Number of (incoming/outgoing)  edges(v) |
\end{equation}

\subsection{Closeness Centrality}
\label{S:2}
Closeness centrality is another measurement to determine the importance of vertices on a global scale within a graph. In this scenario, the closeness of each node to all other nodes in the graph is calculated as a metric to show the importance of each node. A node is usually considered as a key node 
if it can quickly interact with all the other nodes in a graph, not only with the first neighbors. In the literature the closeness	is	defined  as the	length	of	the	average	shortest	path	between	a	vertex	and	
all	vertices in	the	graph\cite{sabidussi1966centrality,freeman1977set}.	The $Closeness Centrality$ of node $v$ is calculated in equation  \ref{eq:cc} where $d(v,j)$ is the minimum number of edges to get from node $v$ to node$u$ 
\begin{equation}
\label{eq:cc}
CC(v) =  \dfrac{ N-1 }{ {\sum\limits_{}{_{u \in G} d(v,u)  }}  }
\end{equation}

\subsection{Betweenness Centrality}
\label{S:2}
Betweenness centrality which is primarily focused on the position of a vertex in a graph is defined as the number of shortest path from all nodes in a graph to all other nodes that pass through a particular node. The betweenness centrality which was originally proposed in \cite{freeman1977set} concentrates on undirected and unweighted graph. This measure is generalized for directed graphs in \cite{white1994betweenness} and weighted directed graphs in \cite{brandes2001faster}. The same as degree centrality, a node with a higher betweenness value is considered more important.
\begin{equation}
\label{eq:emc}
BC(v) = \sum\limits_{s\neq v\neq t} \dfrac{ \sigma_{st}(v)  }{ \sigma_{st}  }
\end{equation}
Where $\sigma_{st}$ is the total number of shortest paths from node $s$ to node $t$ and $\sigma_{st}(v)$ is the total number of those paths pass through node $v$.
\subsection{Bridging Centrality}
\label{S:2}
Information flow and topological locality of a node is calculated through Bridging Centrality measure. Usually, a node connecting densely connected components in a graph is recognized as a node with higher Bridging Centrality value. Bridging Centrality is based on two key factors including betweenness centrality($BC (v)$) and bridging coefficient ($Br_{coefficient} (v)$) of each node.
\begin{equation}
\label{eq:emc}
BrC(v) = Br_{coefficient} (v) . BC (v)
\end{equation}
\begin{equation}
\label{eq:emc}
Br_{coefficient}(v) =  \dfrac{ DC(v)^{-1}  }{ {\sum\limits_{}{_{i \in N(v). \dfrac {1} {DC(i)}}  }}  }
\end{equation}
Where $DC(v)$ is the degree of node $v$ and $N(v)$ is the set of it's neighbors.
\subsection{Harmonic Centrality}
\label{S:2}
The modified version of \emph{Closeness} approach in a graph  \cite{boldi2014axioms} is named harmonic centrality in which the average distance is replacing with the harmonic mean of all distances. 
\begin{equation}
\label{eq:emc}
HC(v) =  \dfrac{ 1  }{ {\sum\limits_{u\neq v}{{d(u,v) }  }}  }
\end{equation}
\subsection{Radiality}
\label{S:2}
Radiality calculates the closeness of a particular node to all nodes in a graph through computing the diameter of a graph\cite{valente1998integration}.
\begin{equation}
\label{eq:emc}
Ra(v) =  \dfrac{ 1  }{ {\sum\limits_{u\neq v}{{ \bigtriangleup_G -(1/d(u,v))      }  }}  }
\end{equation}
where $\bigtriangleup_G$ is the  diameter of a \emph{G} 
\subsection{Ego Centrality}
\label{S:2}
Ego Centrality measurement for a node \emph{v} aims to generate a subgraph of the main graph \emph{G} including node \emph{v} and its neighbors and all of the edges between them. Defining the importance of node \emph{v} to its neighborhood is the ultimate goal of this technique.

\begin{equation}
\label{eq:emc}
EC(v) = 
 \sum\limits_{i = 1}^{i=n^{in}}{W_i*e.ego_i} + \sum\limits_{i = 1}^{i=n^{out}}{W_i*e.ego_i}
\end{equation}
where: 
\begin{equation}
\label{eq:emc}
W_i = 
 \sum\limits_{i = 1}^{i=n^{in}}{\dfrac{1}{v_i^{out}}} +  \sum\limits_{i = 1}^{i=n^{out}}{\dfrac{1}{v_i^{in}}}
\end{equation}
where \emph{e.ego=1/$v_i^{out}$}, \emph{$v_i$} the adjacent node of a node \emph{v} using the incoming edge \emph{e} and \emph{e.ego=1/$v_i^{in}$}, \emph{$v_i$} the adjacent node of a node \emph{v} using the outgoing edge \emph{e} 
\subsection{Relative Cardinality}
\label{S:2}
The cardinality of a node in a schema graph is the number of instances in data graph corresponding to that node in the schema graph \cite{troullinou2017ontology}. A node with a higher corresponding instances is expected to be more important compare to a node with a lower instances. The relative cardinality measure can be applied to an edge in a schema graph. In this scenario, the cardinality of an edge between two nodes is calculated as the number of corresponding instances to those nodes in schema graph with that specific edge.
\subsection{Eigenvector Centrality }
\label{S:2}
Usually, nodes with more edges are recognized to be more important nodes in a graph. However, in real-world scenarios, sometimes the importance of the neighbor nodes is a key point in which more important
neighbors provides a stronger signal in comparison with quantity of neighbors \cite{zafarani2014social}.
In fact, eigenvector centrality, can be considered as a degree centrality measurement while we try to incorporate the importance of the neighbors.
The eigenvector centrality of node $v$  in a graph $G$ is calculated as a proportional function of the summation of its neighbors’ centralities \cite{zafarani2014social}.

\begin{equation}
\label{eq:emc}
EiC(v_i) =\dfrac{ 1 }{ {\lambda   }  }\sum\limits_{j = 1}^{n}{A_{j,i}EiC(v_j)} 
\end{equation}
Where $A$ is adjacency matrix of a graph $G$ and $\lambda$ is some fixed constant.


\subsection{Frequency }
\label{S:2}
 Frequency measurement is usually applicable for the cases that we can obtain the main ontology through merging several local ontologies. Ontology merging is the process of combining (merging) two or more local ontology in order to reach to one target ontology \cite{noy2000algorithm}. In \cite{pires2010summarizing}, they came with this assumption that ontology \emph{ (O)} is a merged ontology from local ontologies \emph{($O_1$,..., $O_n$)} and the concept \emph{C} correspond to one or more concepts contained in \emph{($O_1$,..., $O_n$)}. The frequency of concept \emph{C} is calculated via equation \ref{eq:fr}.
\begin{equation}
\label{eq:fr}
Fr(C) =\dfrac{ |Correspondences(c)| }{ { |O_1,..., O_n|   }  }  
\end{equation}
Where\emph{$|Correspondences(c)|$} is the number of concept correspondences involving \emph{c} and \emph{$|O_1,..., O_n|$} is the number of distinct local ontologies.

\subsection{Name Simplicity }
\label{S:2}
Name simplicity \cite{peroni2008identifying} which is originally inspired from \cite{rosch1978principles} under notion of \emph{natural categories} emphasizes that people characterize the world primarily in terms of \emph{basic objects} rather than more \emph{abstract concepts} and it is a useful basis to recognize good representers of an ontology. Name simplicity measurement of a concept \emph{c}, as we may expect, penalizes concepts consists of  compound words while it favors the concepts with a simple name or label. The name simplicity score of a concept is equal to 1 if the name or the label of that concept would be limited to one word. In the case of compound word the name simplicity score is calculated through equation \ref{eq:names}
\begin{equation}
\label{eq:names}
Name Simplicity(c) =1 - \alpha(nc-1)
\end{equation}
Where $\alpha$ is a constant and  \emph{nc} the number of compounds (words) in the label.
\subsection{Density }
\label{S:2}
Peroni et al. in \cite{peroni2008identifying} considered \emph{density} measure as a structuring criteria to be able to highlight the overall organization of an ontology. This measures the richness of a concept within an ontology based on the number of subconcepts, properties and corresponding instances.There are two sub-measures including \emph{global} and \emph{local density} which are calculated as followings:

\begin{equation}
\label{eq:ns}
GlobalDensity(C) =\dfrac{ aGlobalDensity(C) }{ { max(\{\forall N_i \in O \Rightarrow aGlobalDensity(N_i)\} )  }  }  
\end{equation}
\begin{equation}
\label{eq:ns}
\begin{split}
aGlobalDensity(C) =Number of SubClasses(C) * w_S + \\ Number of Properties(C)*w_P + Number of Instances(C)*w_I
\end{split}
\end{equation}
\begin{equation}
\label{eq:ns}
LocalDensity(C) =\dfrac{ GlobalDensity(C) }{ { maxGlobalDensityNearestClasses(C) }  }  
\end{equation}
\begin{equation}
\label{eq:ns}
Density(C) =GlobalDensity(C) * w_G + LocalDensity(C)*w_L 
\end{equation}
Where $w_S, w_P, w_I, w_G,$ and $w_L$ are the constant weights and  based on \cite{peroni2008identifying} practically equal to 0.8, 0.1, 0.1,0.2, and 0.8 respectively.  
\subsection{Coverage }
\label{S:2}
To calculate the coverage of set of concepts ,$\{C_1,...,C_n \}$, within an ontology, we need to define the \emph{Covered} measure of each concept \cite{peroni2008identifying} first via equation \ref{eq:cover}
\begin{equation}
\label{eq:cover}
Covered(C) = C \cup allSubClasses(C) \cup allSuperClasses(C)
\end{equation}
For the main coverage of set of concepts $\{C_1,...,C_n \}$
\begin{equation}
\label{eq:coverge}
Coverage(\{C_1,...,C_n \}) = \dfrac{ |Covered(C_1) \cup ... \cup Covered(C_n)| }{ { |O|   }  }  
\end{equation}
Where $|O| $ is the number of concepts included in ontology $O$.
\subsection{Popularity }
\label{S:2}
\emph{Popularity} measure proposed by  \cite{li2010ontology} aims to identify concepts that are more common in practice. The \emph{Popularity} of concept $C$ is calculated as a normalized number of results returned by \emph{Yahoo} search engine with the concept $C$ as a keyword.
\subsection{Reference }
\label{S:2}
\emph{Reference} measure for a concept $C$ is defined as a normalized number of entities received from Watson Semantic Web search engine\footnote{http://watson.kmi.open.ac.uk/WatsonWUI/} which reference the concept $C$ in an ontology\cite{peroni2008identifying}. 

\section{Ontology Summarization Techniques }
Ontology summarization is usually considered as an effective way to understand an ontology in order to support different tasks such as ontology reusing in ontology construction development. In literatures, ontology summarization is referred to extractive summarization approach in which the important concepts (at the schema layer) and important entities (at the data layer) are extracted and represented as a summary of an ontology. Based on different measurements in the section \ref{assess}, various methods in ontology summarization have been proposed which are highlighting different criterias to generate a summary for an ontology. In the next section, we are primarily focusing on the approaches which are dedicated in \emph{Ontology Summarization} and providing more details about them. Additionally, Table \ref{OntoSum} represents an overview of the models, the measures and the expected outcome for each model.

\begin{table*}[!th]
\caption{Ontology Summarization Models, Measures, and expected outcomes.}
\resizebox{\linewidth}{!}{%
\begin{threeparttable}
\begin{tabular}{l cccccccccccccccc ll}
\toprule
 & \multicolumn{16}{c}{\textbf{Measures}} &  \\
\cmidrule(lr){2-17} 
\textbf{Model}     &  DC\tnote{1}      &
 CC\tnote{2}   & 
 BC\tnote{3}   & 
 BrC\tnote{4}    & 
 HC\tnote{5}   & 
 Ra\tnote{6}   &
 EgC\tnote{7}   & 
 RC\tnote{8}   &
 Re\tnote{9}   &
 EiC\tnote{10}   &
 Fr\tnote{11}   &
 NS\tnote{12}   &
 De\tnote{13}    &
 Co\tnote{14}    &
 Po\tnote{15}    & 
 Ref\tnote{16}    &

\textbf{Outcome}&\textbf{Note}  \\
\midrule
  \cite{pappas2017exploring}  & \cmark{}  & -  & \cmark{}     & \cmark{}  & \cmark{}   &\cmark{}  &  \cmark{}   & - & - & - &  -  &  - & -   & -  &-    &- & sub-schema graph      & Normalize each importance measure and use \emph{adopted importance measure} \\
\hline
\cite{queiroz2013method}  & \cmark{} & \cmark{} & - & - & - &-  &  -  & -     & \cmark{}     & - &   -  &  -&-   & - & -&-   &Sub-schema graph      &Ontology summary size is defined by a user \\
\hline

\cite{zhang2007ontology,zhang2009summarizing}    &\cmark{} & -  & \cmark{}     & -{}  & -{}   &-& -  & -    & -     & \cmark{} &  -  &  -&-   & -            & - &-  &RDF graph       & The weighted version of \emph{PageRank} and \emph{HITS} algorithem were applied.             \\
   \hline

   \cite{troullinou2017ontology}     & \cmark{}     & -  & -     & -  & -   &-              &  -   & \cmark{}     & \cmark{}     & -  &   -   &  - &-    & -             & -  &-   &Sub-schema graph      &             \\
   \hline
   
\cite{pires2010summarizing}     & \cmark      & -  & -     & -  & -   &-              &  -   & -      & \cmark      & -  &  \cmark   &  - &-    & -  & - &-    &Sub-schema graph      &              \\
    \hline
    
\cite{peroni2008identifying}     & -      & -  & -     & -  & -   &- &  -   & - & - &-  &   -   &  \cmark &\cmark    & \cmark& -  &-   &Sub-graph      &             \\
    \hline
    \cite{li2010ontology}     & -     &  - &  -    &  - &  -  & -  &   -  &  -     &  -     &  -  &    -  & -{}&\cmark{}   & \cmark{}  & \cmark{} &\cmark{}  &Key concepts      &
    
     Using \emph{structural}  and \emph{linguistic} features of ontology \\
 \hline
    
    \cite{troullinou2015rdf}     & \cmark     & -             & -    & - & -  &-             &  -  & \cmark     & \cmark     & -             &   -  &  -&-   & \cmark            & - &-  &RDF graph      &-              \\
    \hline
   
    
  \cite{zhang2011degree}     & \cmark     & -             & -    & - & -  &-             &  -  & -     & -     & -             &   -  &  -&-   & -            & - &-  &RDF graph      &-              \\
    
    \hline
    \cite{ereteo2009semantic}     & \cmark     & \cmark             & \cmark    & - & -  &-             &  -  & -     & -     & -             &   -  &  -&-   & -            & - &-  &Sub-graph      &-              \\
   
    \hline
    \cite{ hsi2003ontological}     & \cmark     & \cmark            & \cmark   & - & -  &-             &  -  & -     & -     & \cmark             &   -  &  -&-   & -            & - &-  &Core concepts      &-              \\
    
     \hline
    \cite{ ereteo2008state}     & \cmark     & \cmark            & \cmark   & - & -  &-             &  -  & -     & -     & -             &   -  &  -&\cmark   & -            & - &-  &-      &-              \\
   
     \hline
    \cite{ zouaq2011towards}     & \cmark     & -            & \cmark   & - & -  &-             &  -  & -     & -     & \cmark             &   -  &  -&\cmark   & -            & - &-  &Concept maps      &-              \\
   
 \hline
    \cite{ hasan2014generating}   & \cmark     & -            & -   & - & -  &-             &  -  & -     & -     & -             &   -  &  -&-   & -            & - &-  &Sub-graph      &-              \\
    
    \hline
    \cite{ graves2008method}    & \cmark     & \cmark            & -   & - & -  &-             &  -  & -     & -     & -             &   -  &  -&-   & -            & - &-  &RDF graph      &-              \\
    
\bottomrule
\end{tabular}
 \begin{tablenotes}
 
    \item[1] Degree Centrality.
    \item[2] Closeness Centrality.
    \item[3] Betweenness Centrality.
      \item[4] Bridging Centrality.
       \item[5] Harmonic Centrality.
        \item[6] Radiality.
         \item[7] Ego Centrality.
          \item[8] Relative Cardinality.
          \item[9] Relevance.
           \item[10] Eigenvector Centrality.
            \item[11] Frequency.
             \item[12] Name Simplicity.
              \item[13] Density.
               \item[14] Coverage.
                \item[15] Popularity.
                 \item[16] Reference.
             
  \end{tablenotes}
\end{threeparttable}
}

\label{OntoSum}
\end{table*}

\subsection{RDF sentence based approach  }
In \cite{zhang2007ontology, zhang2009summarizing}, an RDF sentence is considered as a basic building block in generating ontology summary. The proposed model consists of four main components including \emph{RDF sentence Builder}, \emph{Graph Builder}, \emph{Salience Assessor}, and \emph{Re-ranker}. The key point in this approach is that the user preference will be discussed to determine the weight of links between RDF sentences. In fact,  from \emph{RDF sentence Builder} and \emph{Graph Builder} components the ontology is mapped to a set of RDF sentences and an RDF Sentence Graph is build based on set of RDF sentences and user's preference. The \emph{Salience Assessor} component is responsible to do link analysis on RDF Sentence Graph, generated from the previous component, in order to assess the salience of RDF sentences. This component applies \emph{Degree Centrality, Betweenness Centrality, and Eigenvector Centrality} to assess the RDF sentences and finally, rank them according to their salience. \emph{Re-ranker} component in the last step generate the final summary of the ontology. The coherence of the summary and its coverage on the original ontology are also considered in this section in addition to user-specified salient RDF sentences.
To evaluate the proposed approach the authors used Kendall's tau Statistic \cite{sheskin2003handbook} to measure the agreement between the model's output and human generated results.

\subsection{Personalized Ontology Summary  }
Queiroz-Sousa et al. \cite{queiroz2013method} defines two steps in ontology summarization including finding key concepts and select them to generate a summary.  For the first step, identifying key concepts,  they introduce \emph{relevance measurement} which is inspired from two main measurements including Degree Centrality and Closeness Centrality (equation \ref{eq:person}).
\begin{equation}
\label{eq:person}
\emph{relevance(C)}=\beta * \emph{DC(C)} + \alpha * \emph{CC(C)}
\end{equation}
Where $\alpha +\beta=1$. In the second step, they develop \emph{Broaden Relevant Paths} (BRP) algorithm in order to find the best path within an ontology that represents a set of interrelated vertices with higher relevance. The BRP algorithm aims to generate three lists including \emph{PathSet, NodeSet, and AdjacentNodes}. The \emph{PathSet} list stores the best paths generated by the algorithm. The quality of each path is defined through two metrics including \emph{Relevance Coverage} and \emph{Relevance Degree}. The \emph{Relevance Coverage} is determined by the proportion of the sum of vertices' relevance within a path by the sum of relevance of the vertices with in the original graph. \emph{Relevance Degree} assesses the relevance average within a path by the higher value of relevance in the graph. The \emph{NodeSet} covers all vertices ordered by their relevance values and the \emph{AdjacentNodes} includes the vertices that have relationships with the vertices of paths contained in the \emph{PathSet}. The \emph{AdjacentNodes} arrange the vertices based on \emph{Relation Relevance} score and the \emph{Relation Relevance} is a function of the number of relationships among a vertex and the paths contained in  \emph{PathSet}, the sum of relevance values of that vertex in a particular path over the number of vertices contained in the \emph{PathSet} list, and finally, the relevance of that vertex. The ultimate summary in this approach containing the most relevant concepts with respect to all relationships between those concepts while considering the parameters set by the user.

\subsection{Ontology-Based Schemas in PDMS  }
In Pires et al. \cite{pires2010summarizing} model, degree centrality and frequency measurements are two key points to generate a summary for an ontology. They have applied the extended version of degree centrality in which the type of relationships between concepts are considered in addition to number of relations that each concept has. The frequency measure of each concept in this model also determines the importance of each concept and the combination of two measurements is define as \emph{relevance} score of that concept.
\begin{equation}
\label{eq:emc}
\emph{relevance(C)}=\lambda.\emph{DC(C)} + \beta * \emph{Fr(C)}
\end{equation}
Where $\lambda +\beta=1$. The relevance score of each concept needs to be greater than or equal to relevance score threshold to be considered as a good candidate for the final summary. Finding group adjacent relevant concepts and identifying paths between those groups of concepts are two phases after assigning a relevance score to each concept.

\subsection{Ontology Summarization: An Analysis and An Evaluation  }
Li et al. \cite{li2010ontology}, highlights the lack of consensus in ontology summarization area and try to come up with a generalized approach for ontology summarization while focusing more on facilitating user understanding of ontology using a few space as possible. They mainly concentrate on \emph{linguistic} and \emph{structural} aspects on ontology as the primary features to be looked in ontology summarization.
In light of \emph{linguistic} aspects, \emph{popularity} and \emph{name simplicity} are two criteria to be discussed and \emph{density} and \emph{reference} are two other criteria that need to be considered with respect to \emph{structural} aspects on ontology. For their evaluation they used Kendall's tau Statistic \cite{sheskin2003handbook} which is often applied to calculate the agreements between two measured quantities. 

\subsection{Identifying Key Concepts within an Ontology  }
Automatically identifying the key concepts within an ontology applying topological and lexical criteria including \emph{Name Simplicity, Density, Coverage,} and \emph{Popularity}  is the main idea behind \cite{pires2010summarizing} approach. The ultimate goal in Pires et al. \cite{pires2010summarizing} model is to return a subset of concept from an ontology that match as much as possible to those concepts produced by human experts. This model is focusing more on returning key concepts within an ontology without pay that much attention to generating a graph of extracted important concepts.
\subsection{Ontology Understanding without Tears  }
Troullinou et al. in \cite{troullinou2017ontology} proposed an advanced version of \emph{RDF Digest: Efficient summarization of RDF/S KBs} \cite{troullinou2015rdf} as a new automatically high quality RDF/S Knowledge Bases summary producer. Finding the most representative concepts within schema graph considering the corresponding instances is the key point in generating the summary for an ontology. In this context, the \emph{structure} of the graph and \emph{semantics} of KB are playing important roles in the final summary. In the proposed approach, the importance of each node (concept) is determined through the \emph{Relative Cardinality}, in the next step the centrality of each node in the KB is estimated by combining the \emph{Relative Cardinality} with the type and number of the \emph{incoming } and \emph{outgoing}edges in the schema. The final step in this approach is generating valid sub-schema graphs that cover more relevant nodes by minimizing their overlaps. Two algorithms that try to optimize the local and global importance of the selected paths are applied in order to generate the final summarized sub-schema graphs.
\section{Conclusions}
In this paper, we investigate different ontology summarization measures and models. While there are automatic methods to generate ontology summaries based on different requirements and tasks but there is still a room to extract more reliable summaries. To best of our knowledge, the procedure of extracting summaries in the current methods is static which means that the summaries are produced based on some pre-defined measures. In an ideal case, the ontology summarization technique needs to be more flexible in the way that users or applications be able to tune the model in order to generate different summaries based on different requirements. 

The available approaches apply \emph{extractive technique} in order to generate the final summary (the exact nodes from the original ontology are selected as a summary). \emph{non-extractive} ontology summarization is a new direction in this area which can be applicable in various applications such as ontology tagging.

\nocite{ li2010evaluations,maedche2002measuring}
\bibliographystyle{IEEEtran}
\bibliography{mybib}

\end{document}